\documentstyle[12pt,epsfig,worldsci]{article}
\begin{document}
\vspace*{-1.3cm}
\begin{flushright}
FERMILAB-CONF-93/224 \\
OCIP/C-93-10\\
hep-ph 9308281
\end{flushright}
\vspace{-0.7cm}
\title{Single Top Production in $e\gamma$ Collisions
\thanks{Talk presented by E. Yehudai at The 2nd International Workshop
on Physics and Experiment with Linear $e^+e^-$ Colliders, Waikoloa,
Hawaii, Apr. 26-30, 1993.}}
\author{
Eran Yehudai \\
{\em Theoretical Physics Group, M.S. \#106 \\
 Fermi National Accelarator Laboratory  \\
P.O. Box 500, Batavia, IL 60510} \\
\medskip
 and \\
\medskip
Stephen Godfrey and K. Andrew Peterson \\
{\em Ottawa Carleton Institute for Physics, \\
 Department of Physics, Carleton University, \\
Ottawa, Ontario CANADA K1S 5B6}
}

\maketitle

\setlength{\baselineskip}{2.6ex}

\begin{center}
\parbox{13.0cm}
{\begin{center} ABSTRACT \end{center}
{\small \hspace*{0.3cm}

We study single top production in high energy $e\gamma$ collisions.
The process $e\gamma\to\nu b\bar t$ can serve as a unique tool
for measuring the $Wtb$ coupling. We show that allowing for realistic
luminosity and backgrounds, the size of this coupling can be
determined to within 10 (5)\% in an $e\gamma$ collider built around a
500 (1000) GeV $e^+e^-$ collider.}}
\end{center}

The top quark mass can be accurately determined in a high energy
$e^+e^-$ collider, using either continuum top decay or threshold top
production.\cite{mass}
The top width, however, is much more difficult to measure. In the
Standard Model, this width is proportional to the $Wtb$ coupling
which, in turn, is proportional to the KM element $|V_{tb}|^2$.
If $|V_{tb}|$ is found to be significantly less than 1, it could serve
as evidence for mixing with a fourth generation of quarks.

It has been proposed\cite{jikia} that the process $e\gamma\to
b\bar t$ could be used to directly measure $V_{tb}$. The total cross
section for the process is directly proportional to $|V_{tb}|^2$. In
this work we examine this claim in detail, by considering realistic
luminosity, cuts and backgrounds.

The Feynman diagrams contributing to the process $e^-\gamma\to\nu
b\bar t$ are given in Fig.~1. The top quark is not observed directly.
Rather, it decays into a $W$ and a $b$. In looking for appropriate
cuts and potential backgrounds, one has to take into account the
correlation between the top production and subsequent decay.
We chose to do that by including the top decay as part of the Feynman
diagram. By appropriately selecting the $W$ polarization
vectors we included correlations with the subsequent $W\to q\bar q'$
decay.

\begin{figure}
\begin{center}
\mbox{\epsfig{file=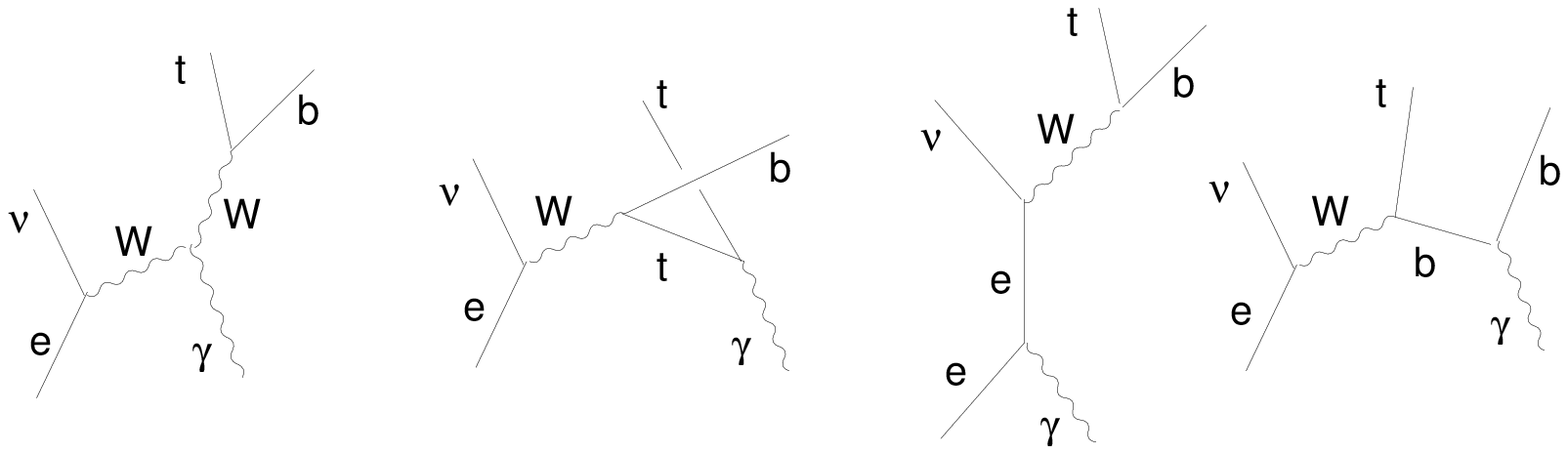,width=6in}}
\end{center}
\noindent
{\small Fig.~1.  The Feynman diagrams contributing to the process $e\gamma \to
\nu b\bar t$. }
\end{figure}

An analytic formula for the differential cross section of
$e\gamma\to\nu b\bar t$ is given in ref.~1. We have used this formula
to check our results by integrating over top decay phase-space.
The actual calculation was carried out using the Vector Equivalence
Technique\cite{ve} and was checked using the helicity amplitude
techniques of Hagiwara and Zeppenfeld\cite{hagiwara}.

The major backgrounds we are concerned with are $e\gamma\to \nu WZ$
and $e\gamma\to e W^+W^-$. Fig.~2 shows the energy dependence of the
total cross sections of the process $e\gamma\to W b\bar t$ and the
various backgrounds\cite{cheung}.

\begin{figure}
\begin{center}
\mbox{\epsfig{file=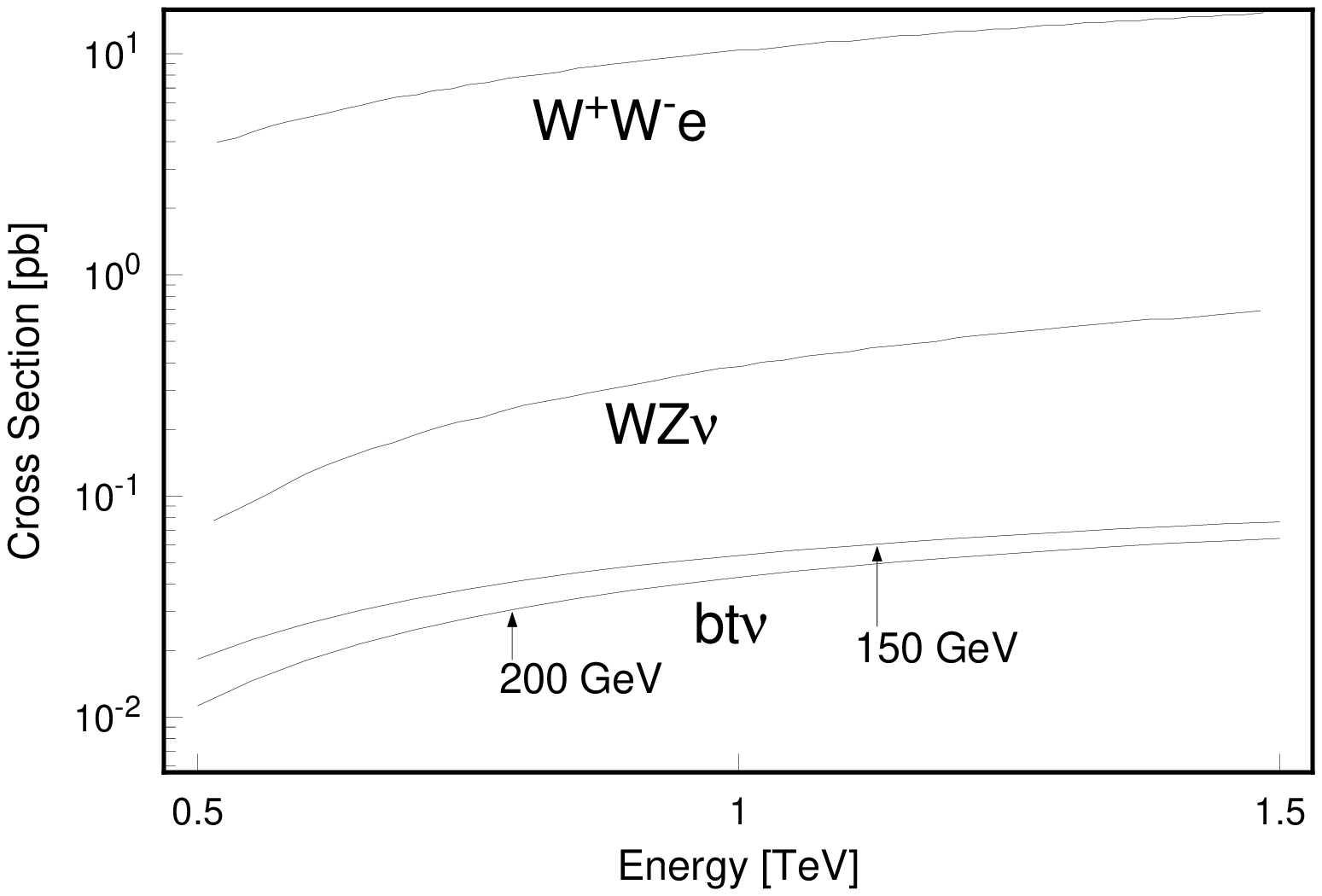,height=2.55in,angle=90}}
\end{center}
\noindent
{\small Fig.~2.  Total cross section for $e\gamma\to \nu b\bar t$ and
its important backgrounds. }
\end{figure}

If the $Z$ or one of the $W$ bosons decays hadronically, its two jet
decay products could look like the two jets coming from the $b$ and
$\bar b$. The most effective method of eliminating these backgrounds
is by imposing a cut on the $b\bar b$ invariant mass $m_{b\bar b}$:
\begin{equation}
\min\left(\left|m_{b\bar b}-m_Z\right|,
\left|m_{b\bar b}-m_W\right|\right)  > 10 {\rm GeV}.
\end{equation}
A small number of hadronically decaying $W$s or $Z$s could have a
measured invariant mass far enough from $m_W$ or $m_Z$. A future study
should address the question of whether those could still constitute a
significant background.

In addition to cuts necessary to elliminate backgrounds, we also have
to require that both the top decay products and the $b$ jet are
observed. We assume that no detection is possible less than $10^\circ$
away from the beam pipe. Table 1 presents the effect of these cuts on
the observed signal cross section. The numbers represent the signal
reduction due to each cut separately, as well as the final cross
section when all cuts are imposed. The number of events represents an
integrated luminosity of $10 {\rm fb}^{-1}$ at 500 GeV and $40 {\rm
fb}^{-1}$ at 1 TeV.

\begin{table}[hb]
\begin{tabular}{|l|cc|cc|} \hline
& \multicolumn{2}{c|}{
\vbox to 0.2in {}
$m_t = 150 {\rm GeV}$} &
\multicolumn{2}{c|}{$m_t = 200 {\rm GeV}$} \\
& \multicolumn{2}{c|}{$E_{cm} = 500 {\rm GeV}$} &
\multicolumn{2}{c|}{$E_{cm} = 1000 {\rm GeV}$} \\ \hline
& \hbox to 0.2in{} $\sigma[{\rm fb}]$ \hbox to 0.2in{} &
Events &
\hbox to 0.2in{} $\sigma[{\rm fb}]$ \hbox to 0.2in{} &
Events \\ \hline
Uncut Total & 18.1 & 180 & 42.4 & 1700 \\
 Visible Top Decay Products & 17.0 & 170 & 38.8 & 155 \\
$m_{b\bar b} < 70 {\rm GeV}$ or $m_{b\bar b} > 100 {\rm GeV}$ &
14.4 & 145 & 36.2 & 1450 \\
$b$ Jet Visible & 11.1 & 110 & 22.0 & 880 \\
All Cuts & 8.8 & 90 & 17.4 & 700 \\ \hline
\end{tabular}

{\small Table 1.  Cross Section of $e\gamma\to\nu b\bar t$ with the
various cuts as described in the text.}
\end{table}

It is possible to increase available event sample by relaxing the
demand that the $b$ jet is observable. The experimental signature in
this case would consist of the decay products of a single top, and
nothing else. While this seems like a very distinct signature, it
precludes the possibility of using the $m_{b\bar b}$ cut to eliminate
the backgrounds. Additional study is required to determine the cross
section of relevant background in that case.

As demonstrated by the figures in Table 1, it seems that the total
cross section for a single top production can be determined to within
20\%. Since this cross section is proportional to $|V_{tb}|^2$, it
follows that $V_{tb}$ itself can be determined to within roughly 10\%.

In the case of a 1 TeV collider, the cross section can be determined
to within 10\%, and $V_{tb}$ to within 5\%.

\bibliographystyle{unsrt}

\end{document}